\documentclass{article} 
\usepackage{iclr2022_conference,times}

\usepackage{amsmath,amsfonts,bm}

\def\eqref#1{equation~\ref{#1}}

\def\1{\bm{1}}

\DeclareMathAlphabet{\mathsfit}{\encodingdefault}{\sfdefault}{m}{sl}
\SetMathAlphabet{\mathsfit}{bold}{\encodingdefault}{\sfdefault}{bx}{n}

\usepackage{hyperref}
\usepackage{url}
\usepackage{graphicx}

\title{Learning multi-scale functional representations of proteins from single-cell microscopy data}

\author{Anastasia Razdaibiedina$^*$ \\
Vector Institute for AI Research\\
University of Toronto\\
\texttt{anastasia.razdaibiedina}\\\texttt{@mail.utoronto.ca} \\
\And
Alexander Brechalov$^*$ \\
Terrence Donnelly Centre for Cellular \\ \& Biomolecular Research \\
University of Toronto \\
\texttt{alexander.brechalov@utoronto.ca} \\
}

\iclrfinalcopy 
\begin{document}

\newcommand\nnfootnote[1]{%
  \begin{NoHyper}
  \renewcommand\thefootnote{}\footnote{#1}%
  \addtocounter{footnote}{-1}%
  \end{NoHyper}
}

\maketitle

\nnfootnote{\thefootnote{*} These authors contributed equally to this work.}

\begin{abstract}
Protein function is inherently linked to its localization within the cell, and fluorescent microscopy data is an indispensable resource for learning representations of proteins. Despite major developments in molecular representation learning, extracting functional information from biological images remains a non-trivial computational task. Current state-of-the-art approaches use autoencoder models to learn high-quality features by reconstructing images. However, such methods are prone to capturing noise and imaging artifacts. In this work, we revisit deep learning models used for classifying major subcellular localizations, and evaluate representations extracted from their final layers. We show that simple convolutional networks trained on localization classification can learn protein representations that encapsulate diverse functional information, and significantly outperform autoencoder-based models. We also propose a robust evaluation strategy to assess quality of protein representations across different scales of biological function.

\end{abstract}

\section{Introduction}
Systematic analysis of microscopy data has a great potential for fundamental biology and medicine. High-throughput datasets of cellular images contain versatile information about biological properties of molecules, which can be coupled with machine learning approaches to automate drug screening and facilitate the pre-clinical trials. But despite the rapid development of computer vision algorithms, microscopy image analysis still remains a complex problem that has not been completely solved. Currently, one of the biggest challenges is extracting high-quality molecular and genetic representations that can quantitatively characterize protein, gene or chemical perturbation solely based on the image input \citep{caicedo2017data}. Such representations would condense important functional information from the micrograph into low dimensional numerical vectors, and ignore irrelevant information such as noise, cell positions and imaging artifacts. Subsequently, the learned molecular representations can be used to find groups of proteins with similar functions, compare effects of drugs, and characterize unknown genes or chemicals. Alternatively, it is also possible to further fine-tune the learned representations to tailor a particular application. 

Extracting protein representations from single-cell microscopy data is a long-standing challenge. The first computational approaches for automated representation extraction relied on manually designed features \citep{carpenter2006cellprofiler}. Such features included statistics of pixel intensities, cell sizes, textures, distances of the fluorescently tagged proteins to cell periphery etc. Numerical profiles that combined all such quantitative features were used as image representations, and could be subsequently applied for various downstream applications (e.g. cell cycle classification and outlier detection). However, manually choosing features is a very expensive procedure and can result in high performance variability depending on the features design. 

The next generation of approaches used deep learning algorithms to solve various tasks, ranging from cell segmentation \citep{stringer2021cellpose} and classification \citep{kraus2017automated, qin2021multi} to learning representation profiles with autoencoder networks \citep{lu2019learning, cho2022opencell, kobayashi2021self}. A key property of such algorithms is that relevant features are automatically learned from the images by optimizing the training objective. The pioneering works focused on classification of subcellular compartments from single-cell image data using convolutional neural networks (CNNs). Such methods classified cells between 11 to 22 pre-defined localization categories, and showed significant performance improvement over previous feature extraction tools \citep{kraus2017automated, almagro2017deeploc}. Recently, efforts in protein representation learning were directed towards using protein ID information and performing autoencoder-based image reconstruction to learn high-quality features. In this work we re-visit the localization classifier models, and assessed how much functional information can be learned through such training objective. Interestingly, we find that despite simplicity of the objective function, the learned representations contain rich information about proteins' biological properties, outperforming popular representation learning approaches and enabling discovery of bioprocesses within cellular components and even physically interacting proteins.  

Here, we present a simple and robust approach for protein representation learning based on localization classification. Our contributions are as follows:
\begin{itemize}
    \item we show that CNNs trained as subcellular localization classifiers learn representations that achieve state-of-the-art performance on a variety of functional tasks;
    \item we propose a multi-scale evaluation strategy to assess quality of protein representations across multiple functional aspects;
    \item we demonstrate that the proposed approach significantly outperforms current state-of-the-art method for protein representation learning, and a popular feature extraction technique.
\end{itemize}

\section{Methods}
Our goal is to learn molecular representations from microscopy data that encapsulate information about different aspects of protein functioning. We show that representations learned through localization classification objective contain multi-scale functional information, and can be used to detect bioprocesses, pathways and protein complexes. Among existing subcellular localization classification algorithms, we chose DeepLoc model \citep{kraus2017automated} and used representations extracted from its last hidden layer. Despite the fact that DeepLoc is trained to classify subcellular localizations, instead of the final classification probabilities, we can use its intermediate representation profiles to dissect cellular organization at different levels.

\subsection{Models}
\textbf{DeepLoc} \citep{kraus2017automated} is a deep CNN consisting of eight convolutional layers followed by three fully-connected (FC) layers. DeepLoc is trained to predict localization category (17 classes in total) with cross-entropy as an objective function:
\[ L(\widehat{y}_i, y_i) = - \sum_{i=1}^{N} y_i \log(\widehat{y}_i)  \]
where $\widehat{y}_i$ and $y_i$ are predicted probability and ground truth label of the class $i$ ($N$ classes in total).

Original localization classifier models, such as DeepLoc, were used to obtain single-cell probability vectors across localization classes. In our work, we extract representations from the last hidden layer (last FC layer) of the DeepLoc model. We use single-cell image data during training, and after training we average 512-dimensional single-cell representations from the test set to obtain protein's representation. 

\textbf{Paired Cell Inpainting} \citep{lu2019learning} is an autoencoder-based approach that learns to reconstruct protein's fluorescent signal from its cell's cytoplasmic background, and another cell's fluorescent image. Hence, the network requires protein labels during training (to sample cells), its objective function minimizes a standard pixel-wise mean-squared error loss between the predicted target protein \(\widehat{s}_t\) and the actual target protein \(s_t\) ($h$ and $w$ are pixels across image width and height respectively): 
\[ L(\widehat{s}_t, s_t) = \frac{1}{h\cdot w}\sum_{h,w}(\widehat{s}_{t h,w} - s_{t h,w})^2  \]

\textbf{CellProfiler} \citep{carpenter2006cellprofiler} is a modular feature extraction tool used for high-throughput analysis of biological images. CellProfiler is not based on deep learning algorithms, and uses a set of pre-defined features to quantitatively measure cellular phenotypes from the microscopy data, such as intensity, shape, texture etc. Since some of the CellProfiler features can be repetitive, a common post-processing approach for its representations is projection with Principal Component Analysis (PCA). In our work, we used original CellProfiler representation with 433 individual features, and its PCA projection that explains 99\% variance.

\subsection{Dataset and training}
We performed training on a well-characterized  yeast image collection \citep{huh2003global} of 10 million cells containing proteins marked with fluorescent tags. The collection contains 4,049 different proteins (each marked with a fluorescent label), which can be grouped into 17 major subcellular localization categories. The dataset contains two channels - fluorescent green (to mark protein of interest), and fluorescent red (to mark cytoplasm, constant across all proteins) \citep{chong2015yeast}. The dataset was preprocessed by segmenting individual cells and creating 64x64 pixel crops containing a cell in the center. The raw-pixel intensities were standardized for every crop to a mean of 0 and a variance of 1 (individually for every channel). The dataset was split into train, validation and test sets using 80\%, 10\%, 10\% of the cells of each protein respectively. 

We used random flipping and rotation to augment the training data. For each model, we selected the optimal learning rate among $(1e-3, 6e-4, 3e-4, 1e-4, 5e-5)$. The models were trained for 30 epochs, each model had three different runs. We used early stopping to select the best-performing model based on the minimal validation set loss. All the reported results and illustrations are based on the test set data. To obtain protein-level representation, we average all of its single-cell representations from the test set.

\subsection{Evaluation}
Evaluating the quality of molecular representations requires functional assessment at different scales of molecular interaction. In the case of protein representations, we can evaluate whether similar representations indicate that corresponding proteins are a part of the same subcellular compartment, biological process, or protein complex. For that, we use four different standards that characterize the role of a protein - \textbf{GO Cellular Component}, \textbf{GO Bioprocess} \citep{ashburner2000gene, logie2021gene}, \textbf{KEGG pathways} \citep{kanehisa2000kegg, kanehisa2019toward, kanehisa2021kegg} and \textbf{EMBL protein complexes} \citep{meldal2019complex}. Each of these standards evaluates different level of cellular organization, with cellular components grouping proteins with similar morphologies, and protein complex standard capturing the direct molecular interactions (the most stringent requirement).

If learned representations capture functional information, then protein pairs with a higher correlation between their representations are more likely to be part of the same functional group. To evaluate this property, we use three metrics: mean average precision (mAP), adjusted mutual information (AMI) and F-score (with $\beta = 0.5$). Both mAP and F-score assess representation relevance based on a standard using their pairwise correlation, and F-score puts more emphasis on precision. In addition to pairwise functional evaluation, we use AMI to assess clustering efficacy \citep{romano2014standardized}. AMI evaluates how much ground truth labels and clustering labels (derived from representation profiles) correspond to each other.

\section{Results}

\begin{figure}[ht]
\begin{center}
\includegraphics[width=15cm]{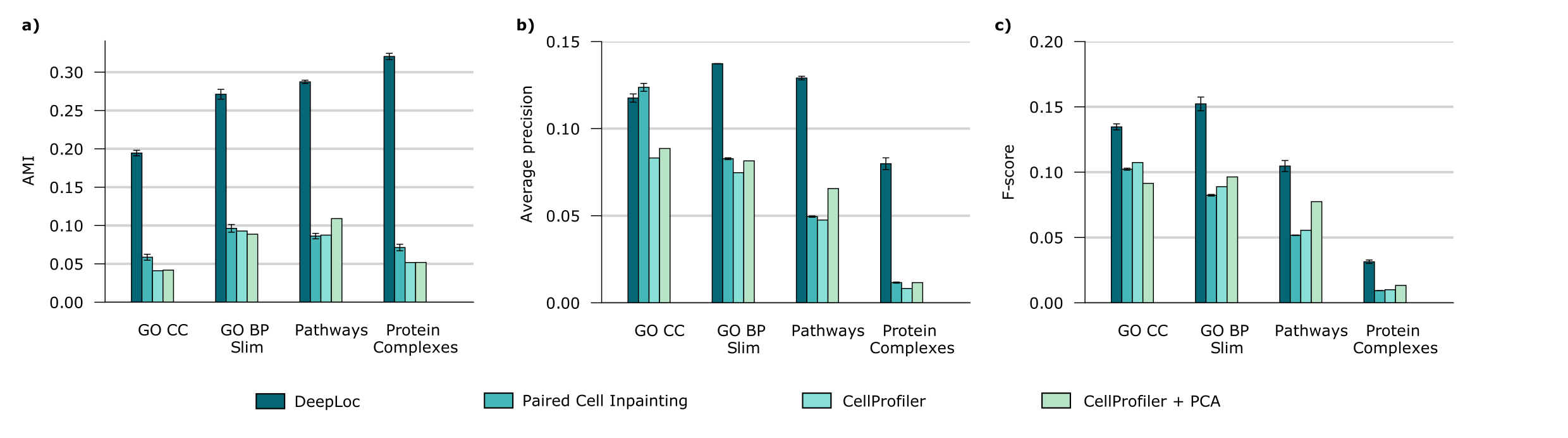}
\end{center}
\caption{Comparison of functional information content of the representations.
Representations obtained from DeepLoc, Paired Cell Inpainting and Cell Profiler were tested for the detecting clusters of functionally related proteins over four biological standards: Gene Ontology (GO) Cellular Component, GO Bioprocess, KEGG pathways and EMBL protein complexes. Performance was assessed with three different scores: a, Adjusted mutual information (AMI score); b, Average precision (AP); c, F-score}
\label{fig:fig1_label}
\end{figure}

\subsection{Methods comparison}
We observe that protein representations learned by DeepLoc encapsulate a variety of functional information. In addition to accurate identification of subcellular components (corresponding to the training objective), representations allow to differentiate between bioprocesses, pathways and even protein complexes (See Figure ~\ref{fig:fig1_label}).

In particular, protein representations derived from DeepLoc achieve 3.7, 2.6, 2.8 and 4.1-fold improvements in terms of AMI over Paired Cell Inpainting on cellular components, bioprocesses, pathways and protein complex standards respectively (Figure ~\ref{fig:fig1_label}a). Similar trends can be observed across the other metrics, AP and F-score. We attribute this to autoencoder-based models being prone to learning noise and imaging artifacts, while robust training objective such as localization prediction ensures that relevant features and morphological information are learned. Surprisingly, representations extracted from DeepLoc can achieve high performance even on standards that are very different from the localization standard that DeepLoc was trained for. This phenomenon has previously been observed in the computer vision community \citep{donahue2014decaf}. DeepLoc representations obtain improvement even on finer standards, such as protein complexes, indicating that by learning to classify localization categories, the model learns to detect comprehensive morphological patterns, yet ignore individual image artefacts.
\begin{figure}[ht]
\begin{center}
\includegraphics[width=11cm]{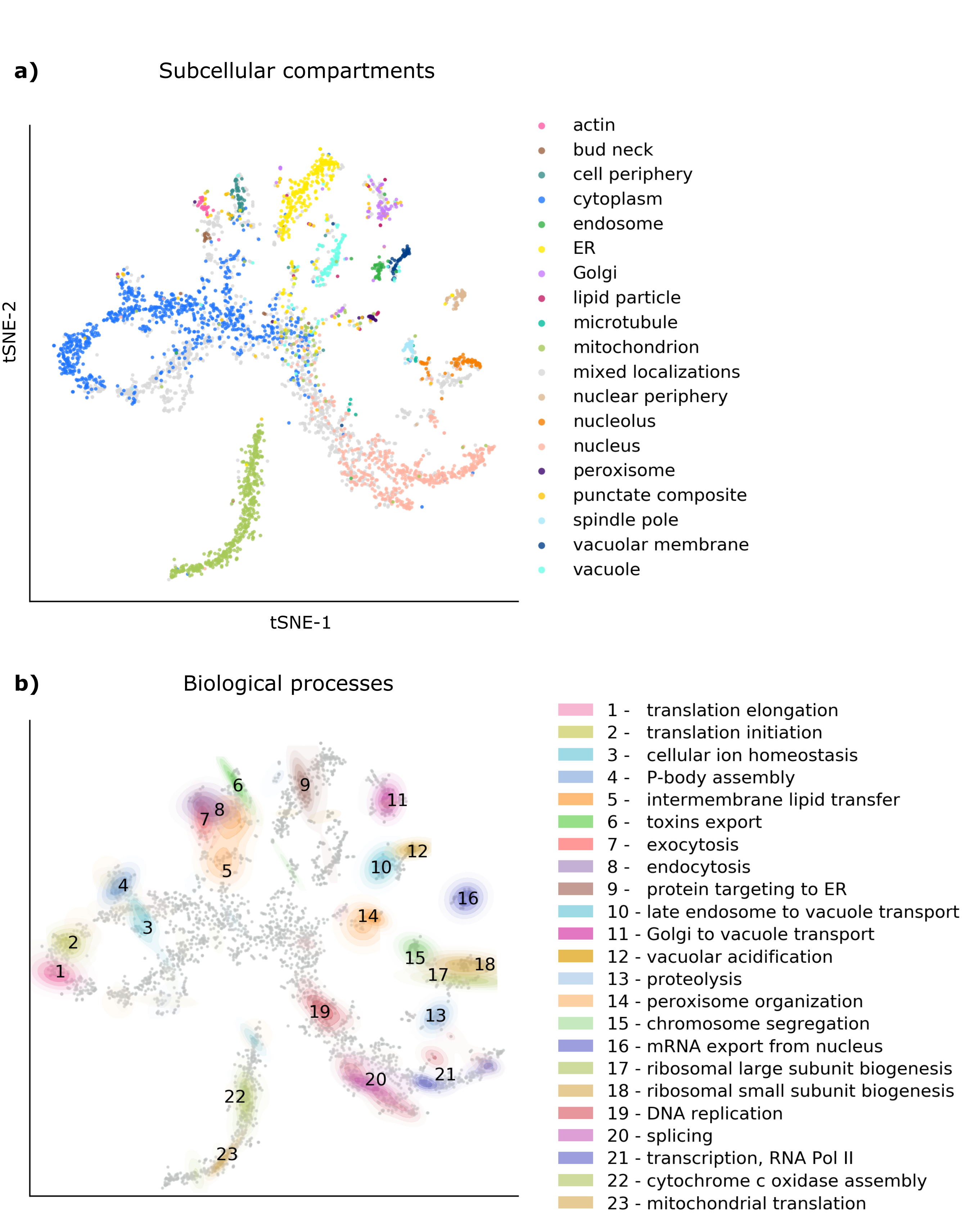}
\end{center}
\caption{Representation profiles of DeepLoc allow to distinguish subcellular compartments and bioprocesses within them.
a, Whole-proteome tSNE projection: each point corresponds to a single protein colored by its localization category. b, Distribution of GO bioprocess categories are shown as Gaussian KDEs , color intensity of the cloud represents cumulative probability mass below it.}
\label{fig:fig2_label}
\end{figure}
\subsection{Visualizing functional representations}
To visualize protein representations learned by DeepLoc, we projected them into 2-dimensional space via t-SNE \citep{van2008visualizing}. Thus, the morphological similarity of proteins encapsulated in the representations was translated into a proximity on the 2D t-SNE map (perplexity = 40) (See Figure ~\ref{fig:fig2_label}a). We showed distribution of 17 localization categories on the top plot. We represented the distribution of 23 fundamental GO bioprocesses on a bottom plot with a kernel-density estimate (KDE) using Gaussian kernels, and used Scott’s rule for KDE bandwidth selection \citep{chiu1991bandwidth}. 

We observe that representations allow to differentiate various biological processes within major cellular components (See Figure ~\ref{fig:fig2_label}b). For instance, translation elongation, translation initiation, P-body assembly and cellular ion homeostasis were discriminated in the cytoplasm. Also, DeepLoc representations allowed to distinguish transcription, splicing and DNA replication processes in the nucleus. Two mitochondrial bioprocesses, mitochondrial translation and cytochrome c-oxydase assembly, were differentiated. Surprisingly, large and small ribosomal subunit biogenesis were distingushed in the nucleolus. Finally, several transport-related bioprocesses (e.g. Golgi to vacuole transport, late endosome to vacuole transport) were also separated.

\subsection{Representation content over the course of training}
\begin{figure}[ht]
\begin{center}
\includegraphics[width=11.5cm]{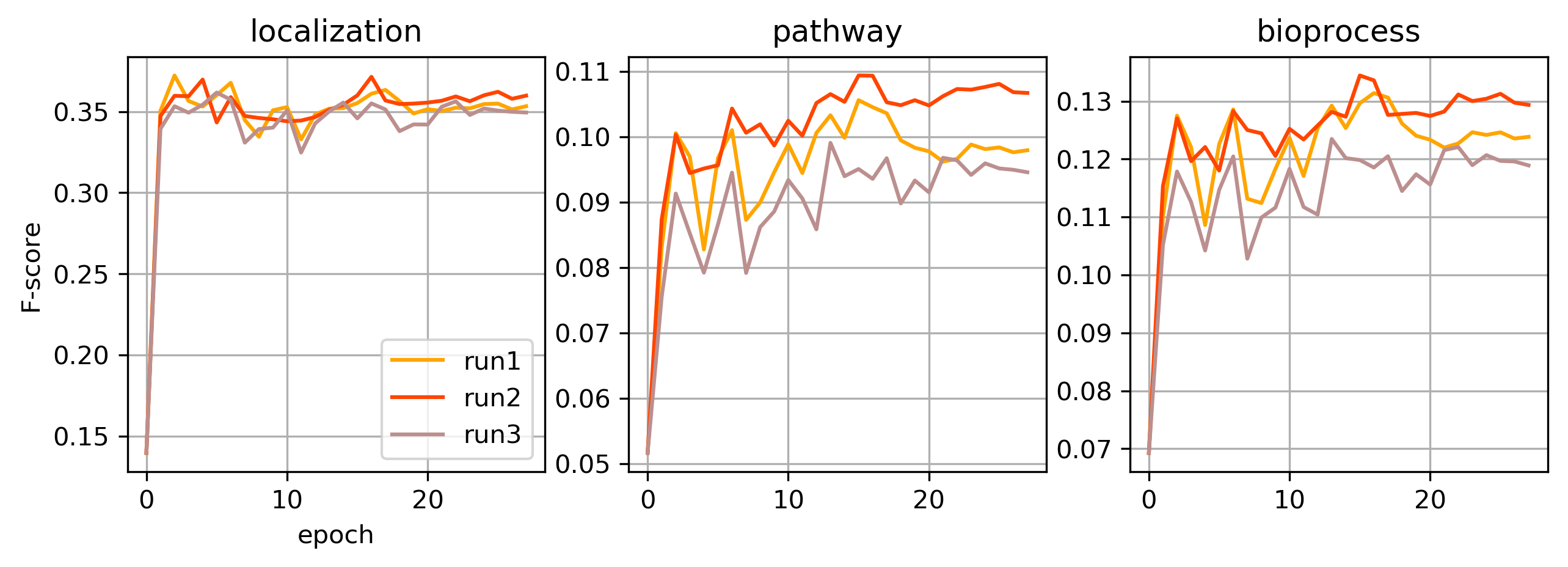}
\end{center}
\caption{Predictive power of the representation profiles over the course of training.
Representation profiles of three random initialization were tested for detecting clusters of functionally related proteins after every training epoch. The performance was assessed with F-score over three biological standards (left to right): GO Cellular Component (localization), KEGG pathways (pathway), GO Bioprocess (bioprocess).}
\label{fig:fig3_label}
\end{figure}

We asked how information content of the protein representations change over the course of training. For that, we monitored F-score of localizations, pathways and bioprocesses detection on test set representations after each epoch (See Figure ~\ref{fig:fig3_label}). We find that the model first learns localization information, according to its objective function, and captures more detailed functional information at later epochs. In particular, F-score on GO Cellular Component prediction reaches its peak after 2nd epoch, while F-score on KEGG pathways and GO Bioprocess keeps increasing until around 15th epoch.

\section{Conclusion}
Learning robust molecular representations from single-cell microscopy data is a challenging task. Current state-of-the-art approaches for protein representation learning use autoencoder-based architectures coupled with reconstruction loss to learn protein-level representation profiles. While such methods put emphasis on learning features relevant to the actual proteins, they are prone to memorizing noise, imaging defects and irrelevant patterns. In this work, we revisited simple convolutional networks for subcellular localization classification, and investigated their use for representation extraction. We find that representations learned through localization classification encapsulate a variety of functional information, allowing to identify bioprocesses, pathways and even groups of physically interacting proteins. Moreover, such representations outperform two common existing methods. Hence, we show that
subcellular localization classification can be used as a strong baseline for protein representation learning.


\subsubsection*{Acknowledgments}
We thank Harsha Garadi Suresh and Helena Friesen for imaging data. We also thank Jimmy Lei Ba for helpful discussions and feedback. We thank anonymous reviewers for valuable
comments and suggestions.

\bibliography{iclr2022_conference}
\bibliographystyle{iclr2022_conference}


\end{document}